\documentclass[preprint,5p,times,twocolumn]{elsarticle}

\usepackage{amssymb}
\usepackage{textcomp,marvosym}
\usepackage[utf8x]{inputenc}
\usepackage{xspace}
\usepackage{url}
\newcommand{\Ldnb}{$\textrm{W}/(\textrm{sr}\cdot\textrm{m}^{2})$\xspace} 

\journal{}

\begin{document}

\begin{frontmatter}
\title{Land use and anisotropy of artificial light observed by night time satellite}

\author{Kai Pong Tong}
\address{Independent researcher; previous affiliation: Institute of Construction
and Architecture, Slovak Academy of Sciences, Bratislava, Slovakia}
\ead{rktong@protonmail.com}

\begin{abstract}
Upward emission of artificial light has been investigated by researchers
since the commissioning of the Visible/Infrared Imaging Radiometer Suite
(VIIRS) Day/Night Band (DNB) in 2011, with applications ranging from night
time light mapping to quantifying socio-economical development. The wide
swath of the VIIRS--DNB sensor enables detection of artificial light at
multiple angles and was utilized to study emission of artificial light from
cities at different angles as well as atmospheric properties. 
Existing studies of the relationship between the directionality and
land surface features are not available for most of the Earth's surface
due to the use of space-borne LiDAR as a source of proxy. To solve this problem,
we compared the land use data published under the Coordination of Information
on the Environment\,(CORINE) against the fit parameters of
radiance of upward artificial light. In general, the quadratic
term of the fit, which quantifies how the brightness changes when viewing
closer from the horizon at a point on the Earth,
is negative when the area is ``Continuous urban fabric'' or 
``Sparsely vegetated areas'', and vice versa
for all other investigated land use classes.
However the quadratic term shifts towards negative values for brighter areas.
These results indicate that
while densely built areas emit more light towards the zenith than sideways,
the VIIRS--DNB is unable to distinguish
small densely built areas scattered around larger unbuilt areas. Therefore,
sensors with higher spatial resolution will be required to
resolve the light emission patterns of areas with
complicated combinations of land uses.
\end{abstract}

\end{frontmatter}

\section{Introduction}
The continued exploitation of various energy sources since more than two
centuries ago, beginning from the industrial revolution\,\cite{owidenergy},
has caused the increased use of artificial light during the night. However,
with the rapid development of solid state lighting device such as the light
emitting diodes (LEDs), the increase in use of artificial light at night is
expected to further accelerate due to
``rebound effect''\,\cite{Tsao_2010,Saunders_2012}, despite their continually
increased energy efficiency and improved LED driver technology which
dramatically increases flexibility in lighting control both in terms of
timing and intensity. In a study conducted in 2017 by Kyba et al., the radiance
of the Earth's surface increased in average by 2.2\% per year and the area
of lit surface, defined by a radiance of $>5\cdot 10^{-5}$\,\Ldnb, increased
by 2.2\% per year\,\cite{Kyba2017_SciAdv}. 

In the past, most measurements of the intensity of emitted artificial light
focused in the direction of zenith (for ground-based measurements) and nadir
(for aerial or satellite-based measurements). In recent years, however,
multi-angle ground-based\,\cite{degen_2022}, aerial and
satellite-based\,\cite{LiX_2019,LiX_2021_multiangle} measurements of
artificial light have attracted scholarly studies due to its impact on the
skyglow pattern, animal behavior and subsequent ecological impact due to
interference on navigation, physiology and predatory patterns. Migratory birds,
for example, are well-known to be severely disoriented when subjected to strong
artificial light sources from long distance in the order of tens of
kilometers\,\cite{VanDoren_2017}. In addition, multiangle measurements can
be potentially used to perform night time optical remote sensing of
atmospheric properties, such as aerosol
loading\,\cite{ZhangJ_2019,ZhangJ_2022_pre}. 

Studies of the blockage effect of buildings for emission of artificial light
at night exist, using either 3-dimensional models\,\cite{Tan_2021}, or with
elevation data from space-borne LiDAR\,\cite{LiX_2021_multiangle}. Efforts were
also made on modeling of the angular emission profile of a city, by
finding out the so-called city emission function (CEF) (see, for example,
Kocifaj et al. 2022\,\cite{Kocifaj_2023a}). However, for most land areas
of the world, these data are not publicly available. In contrast,
land cover data are available for most of the land surface of the Earth,
thanks to multispectral satellite-based remote sensing, which can be freely
obtained by online platforms such as the land use data published under the
Coordination of Information on the Environment\,(CORINE) program of
the European Environment Agency's Copernicus Land Monitoring Service
(hereafter CORINE dataset)\,\cite{corine_2019}.

In this article, we attempt to find out the whether there is any relation
between land use type and the change of artificial light emission towards
space at different viewing angles using the artificial light data of the
Continental Europe obtained by the Suomi National Polar-orbiting Partnership
(S--NPP) Visible/Infrared Imaging Radiometer Suite (VIIRS) Day/Night Band (DNB),
and discuss if land use can serve as a proxy of multiangle artificial
light emission. It should be noted, however, that this analysis can be carried
out in any lit area of the world, as long as there are sufficient data points
for different incidence angles.

\section{Methodology}

\subsection{Night time remote sensing dataset}
The data acquired by the S--NPP VIIRS--DNB sensor for year 2018 was processed
as described in \cite{Tong_2020} (hereafter referred to as
``the Previous Article''). The procedure of the data processing has been
described in detail in the article, and a summary is provided here as follows:
the S--NPP VIIRS--DNB sensor data record (SDR), geolocation data and cloud mask
were downloaded for the designated area of the continental Europe. For each
grid cell, overflight data was selected only when satisfying the following
criteria:
\begin{itemize}
 \item{Sun elevation angle on ground level $ \leq $ --18\textdegree (during
 astronomical night)};
 \item{Moon elevation angle on ground level $\leq $ 0\textdegree (not above
 the horizon)};
 \item{Cloud mask flagged as ``Confidently Clear'' when radiance is below
 $4\cdot 10^{-4}$\,\Ldnb, and ``Probably Clear'' when above, due to performance
 issue of cloud mask in areas of potentially high aerosol load}.
\end{itemize}
After data reduction a quadratic fit is performed on each grid cell. For
consistency, the convention used for the quadratic fit in the Previous Article
is also used here:

\begin{eqnarray}
\label{eq:eq01}
    L_{\textrm{fit}}(\theta) = a\theta^2 + b\theta + L_{\textrm{fit, nadir}}
\end{eqnarray}

where $\theta$ is the AZ-angle (also known as
\emph{directional satellite zenith angle} in other articles such as
Solbrig et al. 2019\,\cite{solbrig_assessing_2020}) $a$ is the quadratic term,
$b$ the slope and  $L_{\textrm{fit, nadir}}$ the fitted nadir radiance.
For the sake of easier comparisons among areas of different radiances and
types of land use, both $a$ and $b$ are divided by $L_{\textrm{fit, nadir}}$:

\begin{eqnarray}
\label{eq:eq02}
 a_{\textrm{rel}} = a / L_{\textrm{fit, nadir}}
\end{eqnarray}
\begin{eqnarray}
\label{eq:eq03}
 b_{\textrm{rel}} = b / L_{\textrm{fit, nadir}}
\end{eqnarray}

where $a_{\textrm{rel}}$ and $b_{\textrm{rel}}$ are the
\emph{relative quadratic term} and the \emph{relative slope}, respectively.

The VIIRS--DNB sensors on board the S--NPP and the other satellites of
the Joint Polar Satellite System (JPSS) continuously acquire data to this date.
However, because the year 2018 is the latest available time period of
the CORINE dataset as of this writing, and because there is also change in
intensity of artificial light over time, in most cases
increasing\,\cite{kyba_2020}, the 2018 S--NPP VIIRS--DNB dataset were chosen. 

\subsection{Land use data source and processing}
The website of the Copernicus Land Monitoring Service publishes the CORINE
land use data both in vector and raster formats, at a resolution of
10\,m and 130\,m respectively. In order to match the resolution of the
processed multiangle VIIRS--DNB data, the raster version of the 2018 dataset
was downloaded and the resolution was reduced to 750\,m, and retained for each
pixel only the most dominant land use.

The CORINE land cover dataset for 2018 was derived from imagery data from the
Sentinel--2 satellite, as well as those from Landsat--8 for the purpose of
gap filling. There are five Level-1, 15 Level-2, and 44 Level-3 classes in the
classification scheme of the CORINE data\,\cite{corine_classes}. Out of the
Level-3 classes, the Classes 111 and 112 are of particular interest, as they
respectively mostly represent dense metropolitan areas / city centers and
suburban areas / villages, and the difference in the light emission pattern
would show qualitatively how the structures of a city / village affects the CEF.

\subsection{Data analysis}
To show the variation of radiance with respect to angle for different use type,
the quadratic fit data were further reduced by selecting only the subdatasets
if there are at least 200 grid cells (corresponding to a constant area of
112\,km², due to the use of the EASE-2.0 Grid, an equal area projection)
with at least 20 overflights in the year and a measured radiance of at least
$5\cdot 10^{-5}$\,\Ldnb. To show the change in these parameters for different
locations of varying radiances, these subdatasets were further divided into
seven bins of equal logarithmic intervals, $\sqrt{3}$-fold apart each,
between $5\cdot 10^{-4}$\,\Ldnb and $2.34\cdot 10^{-3}$\,\Ldnb. For each of
these bins which was selected, the median values, as well as the
16th and 84th percentiles, and the 5th and 95th percentiles (corresponding to
the 1 and 2-standard deviation values, respectively), were extracted.

\section{Results}
\subsection{Relationship between trend of variation in radiance to viewing angle
and land use}
\begin{figure*}[h!]	
\includegraphics[width=18 cm,height=20 cm]{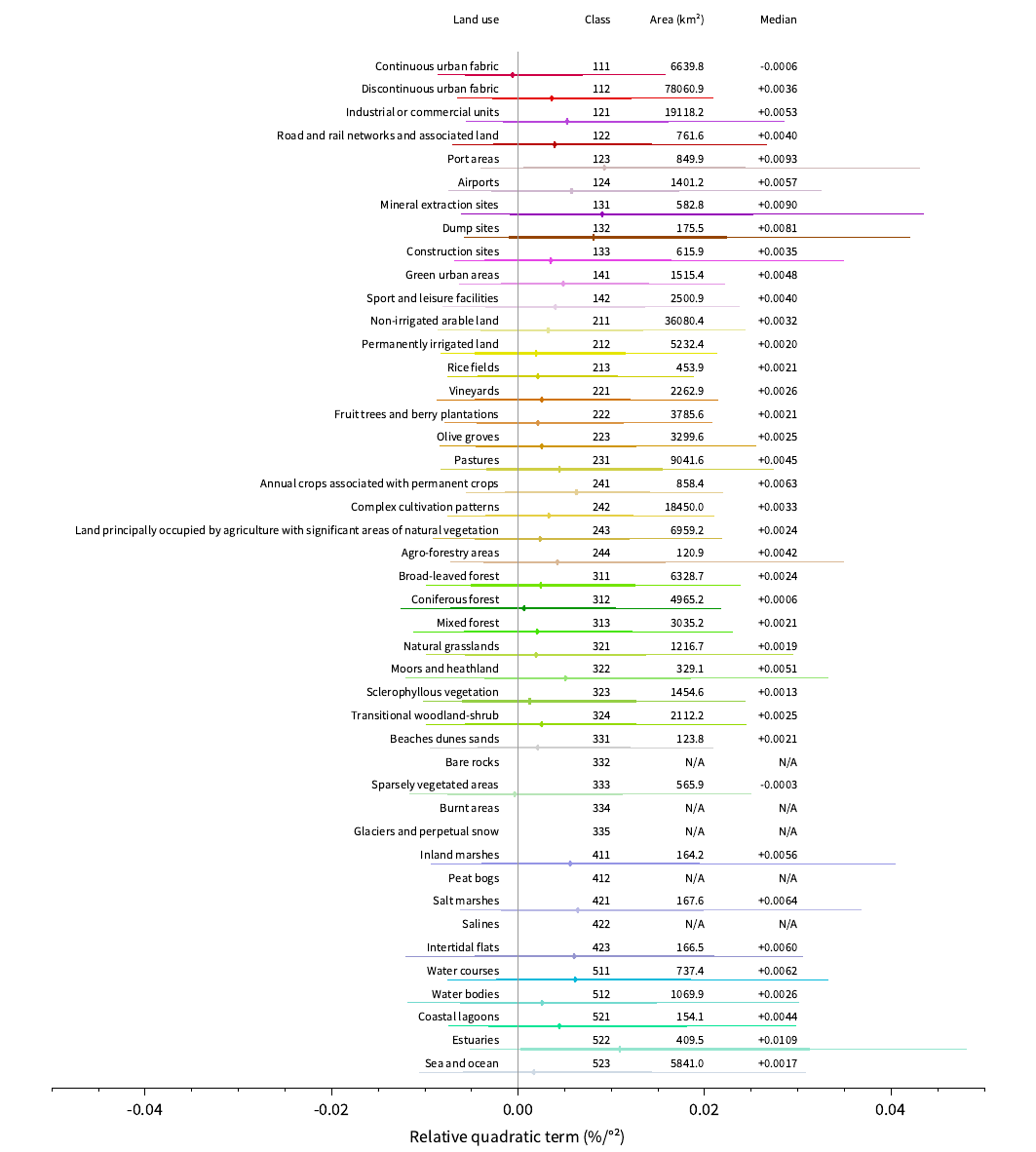}
\caption{Distribution of $a_{\textrm{rel}}$ values for different
land use classes. For each class, the thick error bar represents the interval
between the 16th and 84th percentiles, and for the thin error bar between
5th and 95th percentiles. The vertical bar shows the median value. Only
subdatasets for land use classes where there are at least 200 grid cells
satisfying the selection criteria (at least 20 overflights, no moon or
twilight, fitted nadir radiance at least $5 \cdot 10^{-5}$\,\Ldnb)
are shown in this Figure.\label{fig1}}
\end{figure*}

\begin{figure*}[h!]	
\includegraphics[width=18 cm,height=20 cm]{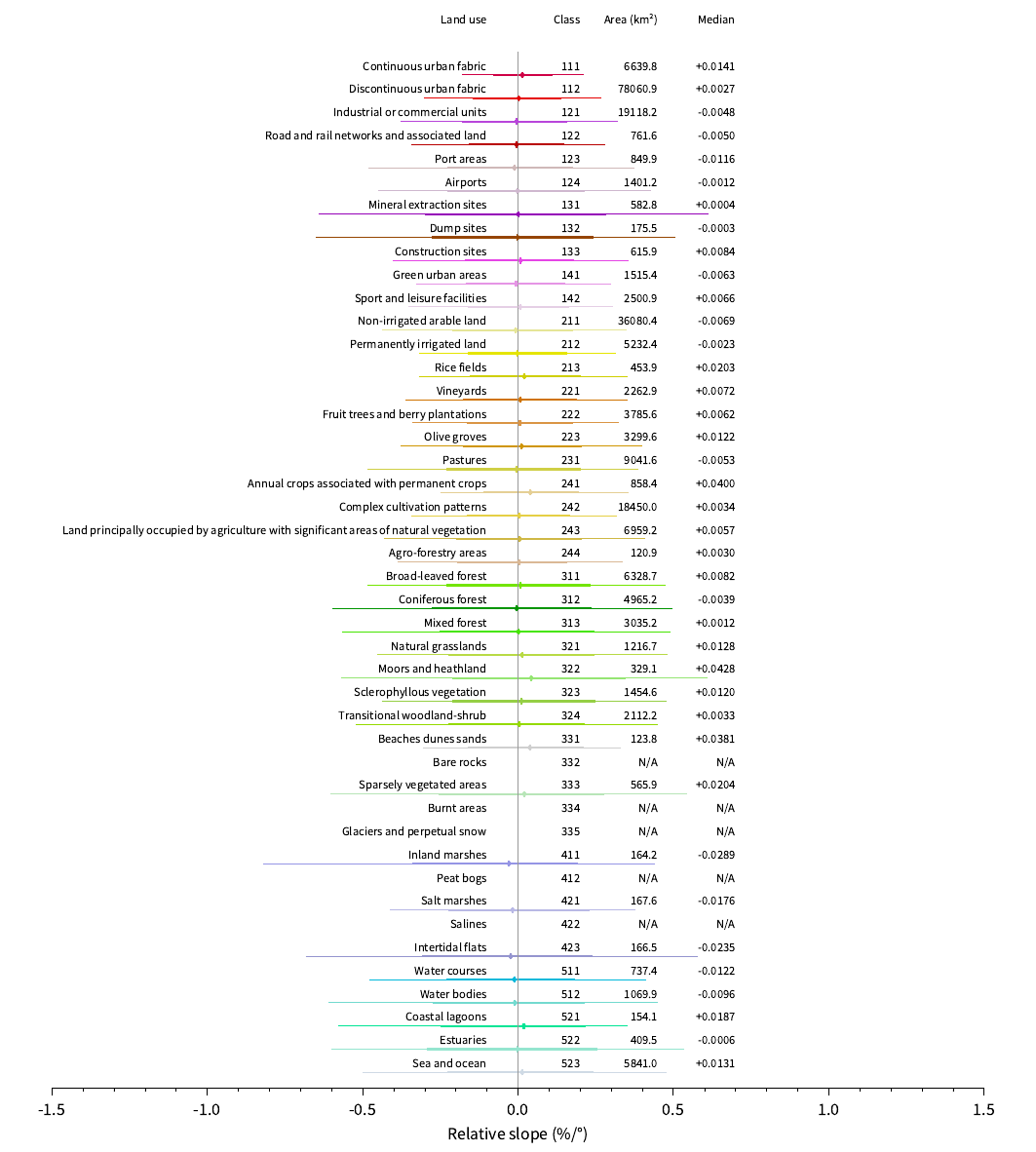}
\caption{Distribution of $b_{\textrm{rel}}$ values for different
land use classes. For each class, the thick error bar represents the interval
between the 16th and 84th percentiles, and for the thin error bar between
5th and 95th percentiles. The vertical bar shows the median value. Only
subdatasets for land use classes where there are at least 200 grid cells
satisfying the selection criteria (at least 20 overflights, no moon or
twilight, fitted nadir radiance at least $5 \cdot 10^{-5}$\,\Ldnb)
are shown in this Figure.\label{fig2}}
\end{figure*}

The error bar plots of the $a_{\textrm{rel}}$ and $b_{\textrm{rel}}$ for
different land uses are shown in Figures \ref{fig1}
and \ref{fig2} respectively.

When considering all lit areas with $L_{\textrm{fit,nadir}}
\ge 5 \cdot 10^{-5}$\,\Ldnb, only two out of the 39 available land use classes
have negative median values of $a_{\textrm{rel}}$, namely ``Continuous urban
fabric'' and ``Sparsely vegetated areas''\,(Classes 111 and 333). Also of note
is that while ``Continuous urban fabric'' and
``Discontinuous urban fabric''\,(Class 112) are of the same Level-2 classes,
which mostly represent densely populated urban areas or city centers, and
sparse suburban or rural areas such as villages, respectively, the two classes
show different trends at higher AZ-angles: more than 50\% of ``Continuous urban
fabric'' areas have $a_{\textrm{rel}} < 0$, which means that the areas emit
more light close to the zenith than to the horizon, and vice versa for
``Discontinuous urban fabric''. No clear pattern can be seen on the relative
slopes amongst different classes of land uses.

\clearpage
\clearpage
\clearpage
\begin{figure*}[h!]	
\includegraphics[width=18 cm,height=20 cm]{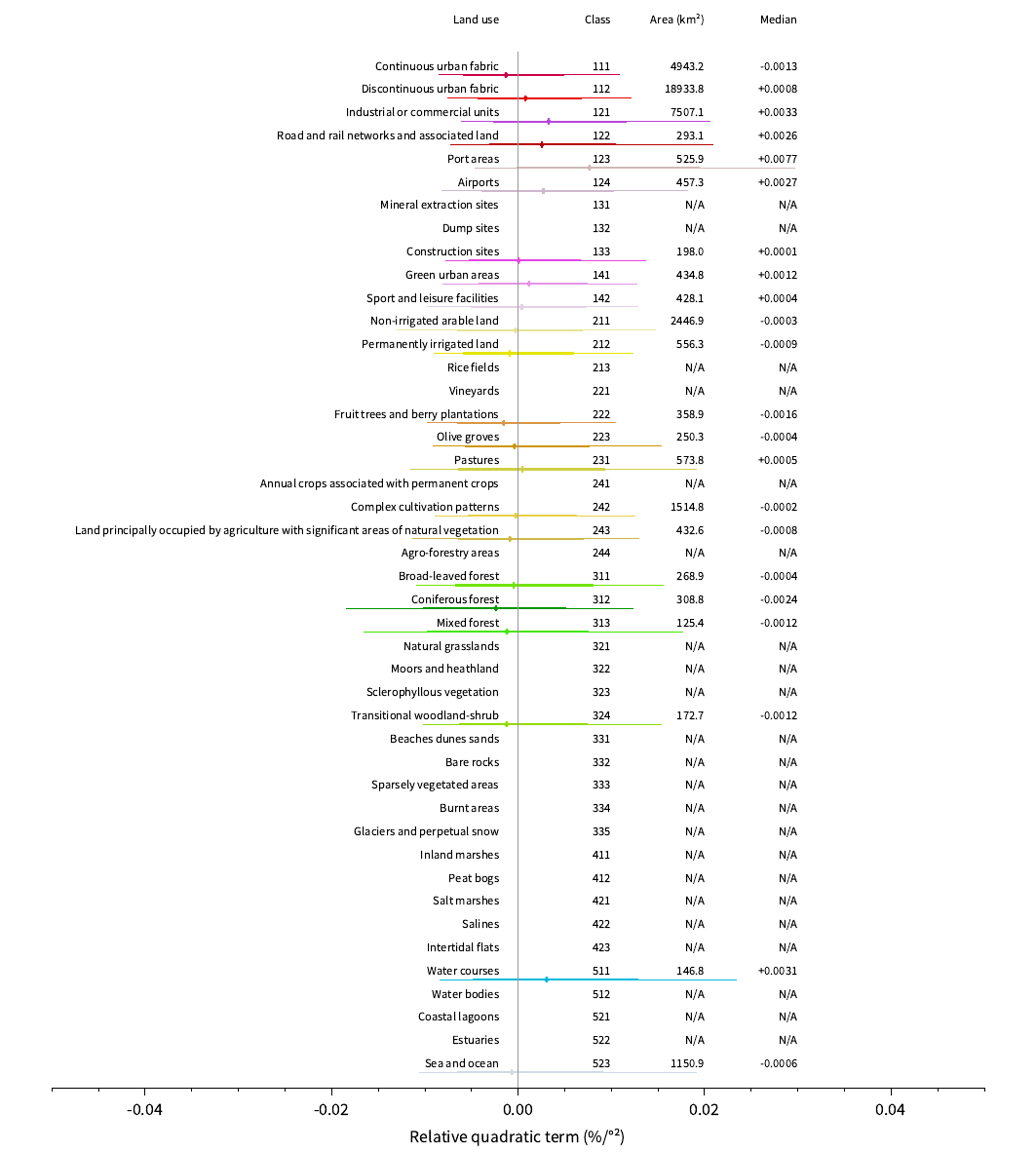}
\caption{Distribution of $a_{\textrm{rel}}$ term values for different
land use classes, using the same plotting scheme as \ref{fig1}, but only for
areas with $2.5 \cdot 10^{-4}$\,\Ldnb $\leq L_{\textrm{fit,nadir}}\leq
1.25 \cdot 10^{-3}$\,\Ldnb).\label{fig3}}
\end{figure*}
\clearpage
\begin{figure*}[h!]	
\includegraphics[width=18 cm]{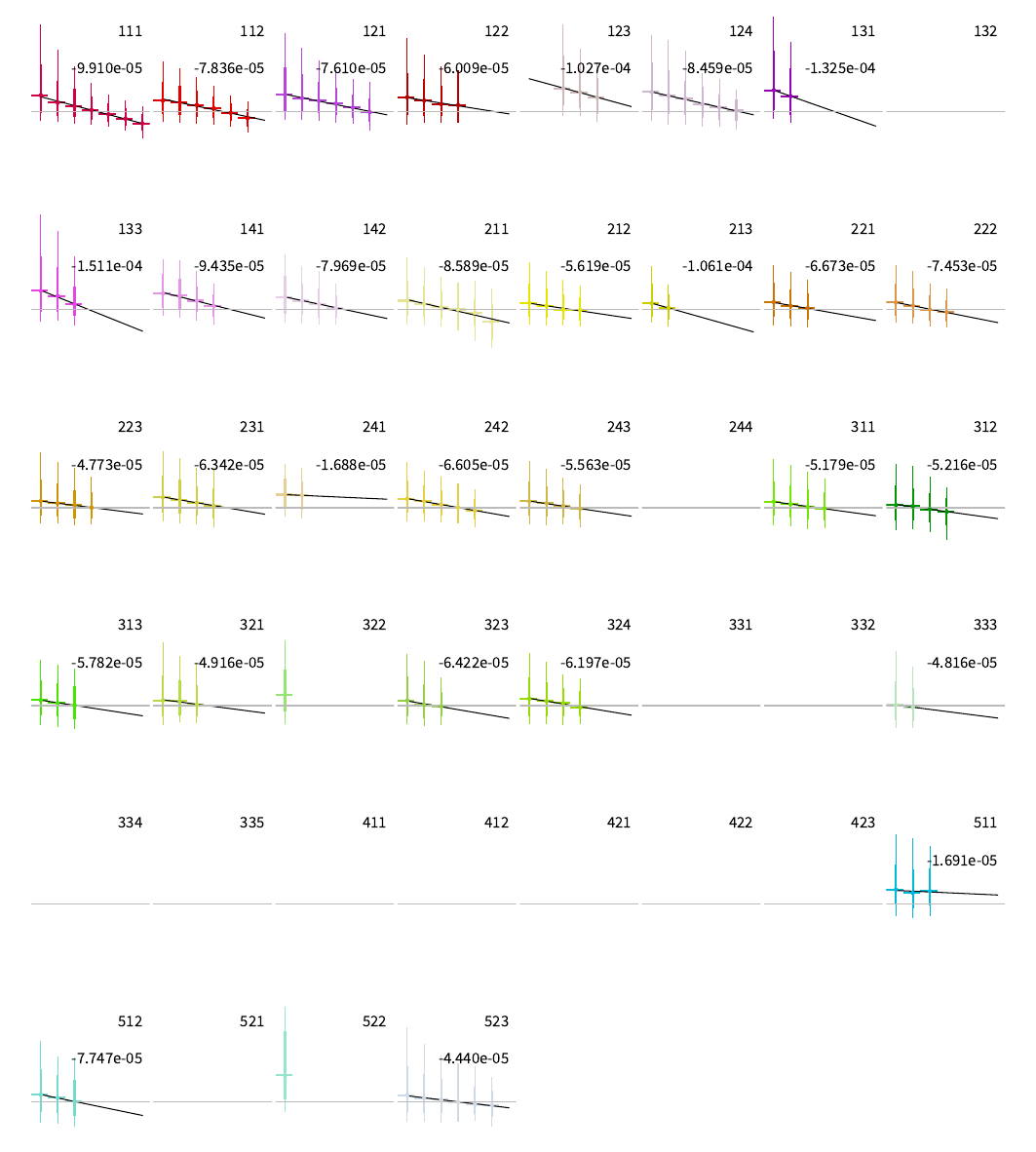}
\caption{Change in distribution of $a_{\textrm{rel}}$ at different
$L_{\textrm{fit,nadir}}$ for land use classes. The x-axis is
$L_\textrm{{fit,nadir}}$ in logarithmic scale binned in the interval of
$\sqrt{3}$-fold each, from $5 \cdot 10^{-5}$\,\Ldnb to
$2.34 \cdot 10^{-3}$\,\Ldnb. For each subplot, the gray horizontal line is
the zero point for $a_{\textrm{rel}}$, the black line the
$log(a_{\textrm{rel}})$ fit with slope of the fit shown in scientific
notation, in the unit of $log(\textrm{sr}\cdot\textrm{m}^{2}/ \textrm{W})
\cdot ^{\circ -2}$, and the class number is shown on the top right corner.
Refer to Figures \ref{fig1} to \ref{fig3} for the corresponding land uses of
the classes.\label{fig4}}
\end{figure*}
\clearpage
\subsection{Emission pattern at different radiance levels}
Figure \ref{fig3} shows the same plot as Figure \ref{fig1}, but with
$L_{\textrm{fit,nadir}}$ between $2.5 \cdot 10^{-4}$\,\Ldnb and
$1.25 \cdot 10^{-3}$\,\Ldnb (five to 25 times of the defined
threshold radiance value). For most classes of land use, the median values
of $b_{\textrm{rel}}$ shift towards the negative side. Figure \ref{fig4}
shows this trend more explicitly: across all land use classes, the higher
the range of $L_{\textrm{fit,nadir}}$ is, the lower the values of
$b_{\textrm{rel}}$. Also of note is that in general, the rate of decrease
in $b_{\textrm{rel}}$ for areas in Level-1 Class 1 is lower than that in the
other Level-1 classes, i.e. while all classes of lands have a decreasing
$b_{\textrm{rel}}$ with increasing $L_{\textrm{fit,nadir}}$, one could
expect that for Class 1 areas, the brighter a particular place is,
the more likely that more light escapes into space sideways
rather than towards the zenith.

\section{Discussion}
As previously reported by multiple literatures, the difference in emission
pattern between urban centers and rural areas can be discerned. More than
half of the areas labeled as ``Continuous urban fabric'' emit more light
towards or close to the zenith than close to the horizon, which is not seen
in almost all other areas in the analysis except one
(``Sparsely vegetated areas'', Class 333). This is in agreement with,
for example, previous findings by Li et al., where LiDAR data from
the United States Geological Survey\,(USGS) were used to investigate
the anisotropy of upward artificial light and found
a moderately strong relationship between several parameters quantifying 
he surface features inside cities, namely the average and standard deviation
of building height, the blocking index of buildings, and the relative change
in radiant intensity with respect to
the viewing angle\,(Figure 10 of \cite{LiX_2021_multiangle}).

When considering the change of the value of $a_{\textrm{rel}}$ with respect to
$L_{\textrm{fit,nadir}}$, there is a negative correlation across different
classes of land uses. This may be due to the fact that at a resolution of
approximately 742\,m, the VIIRS--DNB sensor may not be able to resolve areas
with a high variety of land uses, such as built area close to parks or woods.
This shows the need of satellite remote sensing data from sensors of higher
spatial resolution, especially for densely populated urban areas.

\section{Conclusions}
The ever increasing emission of artificial light at night causes multiple
environmental and social issues, and therefore the interest in
studying patterns of night time artificial light both within and
outside the academia has been increasing in recent years. Although there exist
multiple studies of global artificial light emissions using satellites
capable of wide-area, wide-angle night time imaging such as the S--NPP
and the JPSS series, the anisotropy of upward artificial light emission
was not studied until recently, which found out the relationship between
the angular pattern of light emission and the height and density of obstacles,
which in turn may be related to the extent of urbanization.

Based on the Previous Article, this study aimed to find out whether there is
any relationship between the pattern of artificial light emission and
the land use of any particular area covered by the CORINE land use dataset
for year 2018. It was found from the composite night time light data
of S--NPP VIIRS--DNB that when considering all lit areas
(i.e. where $L_{\textrm{fit,nadir}} \ge 5\cdot 10^{-5}$\,\Ldnb),
all except two out of the 44 land use classes emit more light towards
or close to the zenith than sideways ($a_{\textrm{rel}} > 0$). One of the
land classes where $a_{\textrm{rel} < 0}$ is
Class 111 (Continuous urban fabric), which mostly consists of city centers
or densely populated residential areas. In contrast, Class 222, which mostly
represents more sparsely populated settlement such as villages, exhibits
the opposite pattern. This is in agreement with the previously
satellite-based observations that due to presence of obstacles,
densely populated urban areas emit more light towards or close to the zenith,
and vice versa for suburban/rural areas.

We also found out that when using the VIIRS--DNB sensor, there is a
negative correlation across all land use classes for the term
$a_{\textrm{rel}}$. We suspect that this is due to the sensor's low resolution
relative to the land use dataset, where higher densely populated/built areas
lie within or beside sparsely built/more pristine areas and identified as such.
This shows that sensors with better spatial resolution is needed to resolve
lit areas at street level, or more ideally of individual light sources, which
would require a resolution of approximately 10\,m. In addition, while sensors
with similar capabilities do exist as of this writing, the data are not
licensed in permissive terms, which increases the cost of conducting
similar investigation. More permissive license for the data will open
a new opportunity not only for this particular application, but also
other uses of night time light data products, for example
in social studies\,\cite{kyba_nightwatch_2024}.

\section*{Conflicts of interest}
The author declares no conflicts of interest.

\section*{Data availability}
Data used in this research, which is based on the results of the Previous
Article, is available at

\url{https://doi.org/10.5281/zenodo.16752528}.

\section*{Acknowledgements}
This research was partly funded by the Slovak Research and Development
Agency under contract No. APVV-18–0014. The author would like to thank
Dr. Torben Frost for providing suggestions on the manuscript.

\bibliographystyle{elsarticle-num-names} 
\bibliography{ref}

\end{document}